\documentclass[10pt,letterpaper]{article}
\usepackage{opex3}
\usepackage{mathrsfs}
\usepackage{bm}
\usepackage{amsmath}
\usepackage{cite}
\begin{document}
\title{The role of magnetic dipoles and non-zero-order Bragg waves in metamaterial perfect absorbers}
\author{Yong Zeng,$^{1}$ Hou-Tong Chen,$^{2}$ and Diego A. R. Dalvit$^{1}$}
\address{$^1$Theoretical Division, MS B213, Los Alamos National Laboratory, Los Alamos, New Mexico 87545, USA\\
$^2$MPA-CINT, MS K771, Los Alamos National Laboratory, Los Alamos, New Mexico 87545, USA}
\email{$^{\ast}$yongz@lanl.gov}
\begin{abstract}
We develop a simple treatment of a metamaterial perfect absorber (MPA) based on grating theory. We analytically prove that the condition of MPA requires the existence of two currents, which are nearly out of phase and have almost identical amplitude, akin to a magnetic dipole. Furthermore, we show that non-zero-order Bragg modes within the MPA may consume electromagnetic energy significantly.
\end{abstract}




\section{Introduction}

The experimental demonstration of near-unity absorption in ultra-thin metal-dielectric-metal metamaterial structures \cite{landy}
opens many potential applications in sensing, detection, stealth technology, photovoltaics, and thermovoltaics
\cite{Schuller,carl,Atwater,liun,Mehdi,liu2,Taubert,Aydin,Chihhui,mei,cui,Thomas,Dayal}.
The underlying mechanisms of the so-called metamaterial perfect absorber (MPA) have been addressed using effective medium theory \cite{landy,tao1,liu,tao2}, which approximates the whole metamaterial structure as a homogeneous slab with an effective bulk permittivity and permeability. It was found that the retrieved effective permeability can be described by a Lorentzian function around the designed frequency, indicating the appearance of a magnetic resonance \cite{liu}. This observation was further supported by full-wave simulations which showed the currents in the two metallic layers form a circulating current loop \cite{tao2}.
More recently, an interference-based theory of the MPA has been proposed \cite{zhou,chen}. By approximating thin planar metallic layers
as homogeneous impedance-tuned interfaces between their boundary media \cite{holloway}, the near-zero reflection and transmission can be obtained through the interference and superposition of the multiple reflections and transmissions  \cite{chen2,Shchegolkov}. In agreement with the previous numerical observations, the currents in the two metallic layers predicted by this theory have almost equal amplitude and are nearly out-of-phase.

According to the grating theory \cite{born}, the electric field inside a MPA can be expanded as a superposition of Bragg waves. On the other hand, since both theoretical approaches mentioned above disregard the periodic nature of the metamaterial structure, they thereby restrict to zero-order Bragg waves. In this paper we go beyond this approximation by including all orders of Bragg waves inside the MPA. We analytically show that the condition of MPA requires the existence of two currents within the metamaterial, which are nearly out of phase and have almost identical amplitude. Furthermore, using a combination of analytical and numerical arguments, we show that non-zero-order Bragg waves within the MPA consume electromagnetic energy.

\section{Grating theory for the metamaterial perfect absorber}

Let us consider a metamaterial membrane, surrounded by vacuum, which is periodic in the $xy$ plane. For simplicity, we assume its meta-atoms are arranged in a rectangle lattice with primitive lattice vectors $d_{x}\mathbf{e}_{x}$ and $d_{y}\mathbf{e}_{y}$. Here $d_{x}$ and $d_{y}$ are the corresponding lattice constants. We further assume that the external illumination is a plane wave propagating along the $z$ direction with a wave vector $\mathbf{k}^{i}$. Under this external excitation polarization currents, $\mathbf{J}(\mathbf{r},\omega)=-i\omega\epsilon_{0}[\epsilon(\mathbf{r},\omega)-1]\mathbf{E}(\mathbf{r},\omega)$, will appear inside the metamaterial. Here $\epsilon(\mathbf{r},\omega)$ is the permittivity of the metamaterial, a periodic function of $\mathbf{r}$. Using the free space Green's function, we can express the scattered field in terms of the current
\begin{equation}
\mathbf{E}(\mathbf{r}',\omega)=\sum_{mn}\mathbf{E}_{mn}(\mathbf{r}',\omega)=\sum_{mn}\frac{-\pi Z_{0}e^{i\mathbf{k}_{mn}\cdot\mathbf{r}'}}{d_{x}d_{y}\lambda\kappa_{mn}}\int d\mathbf{r}\mathbf{J}_{mn,\perp}(\mathbf{r},\omega)e^{-i\mathbf{k}_{mn}\cdot\mathbf{r}},
\end{equation}
where the integration is performed over one unit cell, $m$ and $n$ are integers, $Z_{0}=\sqrt{\mu_{0}/\epsilon_{0}}$ is the free-space intrinsic impedance, $\lambda$ is the excitation wavelength, and $k_{0}=2\pi/\lambda$ is the free-space wave number. The wave vector $\mathbf{k}_{mn}$ is defined as $\mathbf{k}_{mn}=\mathbf{k}^{i}_{\|}-\mathbf{g}_{mn}\pm\kappa_{mn}\mathbf{e}_{z}$ with $\kappa_{mn}=\sqrt{k^{2}_{0}-|\mathbf{k}_{mn,\|}|^{2}}$, where $\mathbf{g}_{mn}=(2\pi m/d_{x})\mathbf{e}_{x}+(2\pi n/d_{y})\mathbf{e}_{y}$ are the reciprocal wave vectors, and $\mathbf{k}^{i}_{\|}$ is the projection of the incident wave vector onto the $xy$ plane. Here the positive sign corresponds to forward scattering (propagating along the positive $z$ direction), and the negative sign corresponds to backward scattering (propagating along the negative $z$ direction). Moreover,
$\mathbf{J}_{mn,\perp}=\mathbf{J}-\mathbf{k}_{mn}(\mathbf{k}_{mn}\cdot\mathbf{J})/k^{2}_{0}=-i\omega\epsilon_{0}[\epsilon(\mathbf{r})-1]\left[\mathbf{E}-\mathbf{k}_{mn}(\mathbf{k}_{mn}\cdot\mathbf{E})/k^{2}_{0}\right]$. It is important to emphasize that $\mathbf{J}_{mn,\perp}$ contains information of all Bragg modes through the total electric field $\mathbf{E}$.

Under certain circumstances $\kappa_{mn}$ is imaginary except for $m=n=0$, so that only the zero order waves can survive in the far-field zone,  e.g. at normal incidence and for $\lambda$ bigger than the lattice constants. The forward wave is then given by
\begin{equation}
\mathbf{E}^{f}_{00}(\mathbf{r}',\omega)=\frac{-\pi Z_{0}}{d_{x}d_{y}\lambda k^{i}_{z}}e^{i\mathbf{k}^{i}\cdot\mathbf{r}'}\int d\mathbf{r}\mathbf{J}_{00,\perp}(\mathbf{r},\omega)e^{-i\mathbf{k}^{i}\cdot\mathbf{r}}.
\end{equation}
The backward wave bears a similar expression except that $\mathbf{k}^{i}$ is replaced with $\mathbf{k}_{\|}^{i}-k^{i}_{z}\mathbf{e}_{z}$. This equation suggests that the tangential component of the incident wave vector is conserved.

\subsection{Role of magnetic dipole}

We now apply these equations to a metamaterial perfect absorber. We use a typical MPA proposed in Ref. \cite{liu} (depicted schematically in Fig. 1), which consists of three thin layers. The first layer is a perforated metallic membrane which is referred to as the cross layer, the second layer, called the spacer layer, is filled with a homogeneous dielectric medium with weak absorption, and the last layer is generally a metallic ground plane. Furthermore, the meta-atoms of the cross layer are arranged periodically in a square lattice. The whole structure possesses both $x$ and $y$ mirror symmetries. Under normal incidence $\mathbf{E}^{i}=e^{ik_{0}z'}\mathbf{e}_{x}$, by symmetry $E_{x}$ is an even function of $x$, while $E_{y}$ and $E_{z}$ are odd functions of $x$. Furthermore, the lattice constant $d$ is smaller than the incident wavelength, so that only the zero-order waves survive in the far field. The forward and backward scattered fields in the far-field zone are hence given by
\begin{eqnarray}
&&\mathbf{E}_{00}^{f}(\mathbf{r}',\omega)=-\frac{Z_{0}\mathbf{e}_{x}}{2}e^{ik_{0}z'}\int_{-h/2}^{h/2}g(z,\omega)e^{-ik_{0}z}dz,\\
&&\mathbf{E}_{00}^{b}(\mathbf{r}',\omega)=-\frac{Z_{0}\mathbf{e}_{x}}{2}e^{-ik_{0}z'}\int_{-h/2}^{h/2}g(z,\omega)e^{ik_{0}z}dz.
\end{eqnarray}
Here $h$ is the thickness of the MPA, and the integral function
\begin{equation}
g(z,\omega)=\frac{1}{d^{2}}\int_{-d/2}^{d/2}\int_{-d/2}^{d/2}J_{x}(x,y,z,\omega)dxdy
\label{eq1}
\end{equation}
stands for the current density at a specific $z$ plane. Note that although the forward and backward waves in the far field are determined by the zero-order Bragg mode, the current $J_{x}$ however contains contributions from all orders of Bragg waves within the structure since the quantity $\mathbf{J}_{00,\perp}$ defined in Eq.(2) equals to $\mathbf{J}_{\perp}$, the transverse component of the total current inside the metamaterial. As a direct result, the reflected wave is $\mathbf{E}_{00}^{b}$, and the transmitted wave is $\mathbf{E}^{i}+\mathbf{E}_{00}^{f}$.

\begin{figure}[t]
\centering
\includegraphics[width=0.7\textwidth]{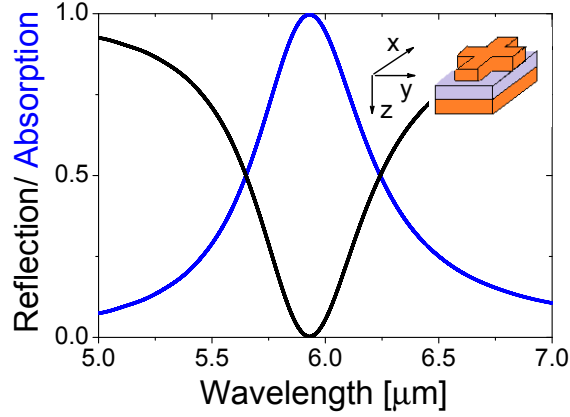}\vspace*{-6.0cm}
\caption{Reflection and absorption spectra of the metamaterial perfect absorber under normal incidence. The inset shows the geometry of the metamaterial. The metallic cross consists of two $0.4\times1.7$ metallic bars. The thicknesses of the cross, spacer and ground layer are 0.1, 0.09 and 0.2, respectively. The total thickness $h$ equals 0.39, and the lattice constant $d=2$. All dimensions are in micrometers.}
\end{figure}

A metamaterial with unity absorption neither reflects nor transmits the incident EM wave to the far field, which is equivalent to requiring
\begin{eqnarray}
&&\int_{-h/2}^{h/2}g(z)e^{ik_{0}z}dz=0,\\
&&\int_{-h/2}^{h/2}g(z)e^{-ik_{0}z}dz=\frac{2}{Z_{0}}.
\end{eqnarray}
To simplify the notation, we dropped the $\omega$ dependence of $g$ here and in the following. The first equation imposes no reflection, and the second one leads to zero transmission. Because the thickness $h$ is much smaller than $\lambda$, we can employ multipolar analysis by expanding $e^{ik_{0}z}$ as $1+ik_{0}z+\cdots$. The leading order gives the electric dipole and the first order of $k_{0}z$ gives the combination of magnetic dipole and electric quadrupole. Supposing these three multipoles dominate the above two integrations, we arrive at
\begin{equation}
\int_{-h/2}^{h/2} g_{R}(z)dz=\frac{1}{Z_{0}}, \:\:\: \int_{-h/2}^{h/2} g_{I}(z)dz=0,
\label{eq4}
\end{equation}
as well as
\begin{equation}
\int_{-h/2}^{h/2} g_{R}(z)zdz=0, \:\:\: k_{0}\int_{-h/2}^{h/2} g_{I}(z)zdz=\frac{1}{Z_{0}},
\label{eq5}
\end{equation}
where $g_{R}$ and $g_{I}$ are the real and imaginary parts of $g$, respectively. Eqs. (\ref{eq4}) immediately suggest that there exists an electric dipole moment with a value of $1/Z_{0}$, and Eqs. (\ref{eq5}) imply that the MPA possesses a magnetic dipole as well as an electric quadrupole, and their sum equals $i/Z_{0}$ (see Eq. (9.31) of Ref. \cite{jackson}). Moreover, as suggested by Eqs. (6) and (7), these three multipoles are destructive along the reflected direction, which leads to zero reflectance.

It is important to notice that the dielectric layer generally has a permittivity $\epsilon_{d}$ which is very different from the permittivity $\epsilon_{m}$ of the metallic layers. For instance, the ratio $(\epsilon_{m}-1)/(\epsilon_{d}-1)$ of the MPA studied below is about $1428e^{i0.94\pi}$ when $\lambda=5.93\:\mu$m. Since
$E_{x}$ is continuous crossing the spacer-ground interface, the polarization current $J_{x}$ is strongly concentrated inside the metallic layers. We therefore can neglect the polarization current within the dielectric layer. Denoting the total current of the cross and ground layer as $G_{c}$ and $G_{g}$, respectively, Eqs. (\ref{eq4}) suggest
\begin{equation}
Z_{0}\times\textrm{Re}(G_{c}+G_{g})=1, \:\:\:\textrm{Im}(G_{c}+G_{g})=0.
\label{eq6}
\end{equation}
In addition, Eqs. (\ref{eq5}) imply $|g_{I}|$ is much bigger than $|g_{R}|$ because $k_{0}|z|\ll1$. We therefore expect that $G_{c}$ and $G_{g}$ are nearly purely imaginary. In other words, the currents in the two metallic layers are nearly out of phase and have almost identical amplitude. This fact, here proved analytically, was reported in earlier numerical simulations \cite{landy,tao1,tao2,liu,Dayal,ma}. It is worth emphasizing that our proof does not require the computation of the current $J_{x}$, which is usually carried out by full-wave simulations.

To support the statements above, we will now perform full-wave simulations of a MPA whose geometrical and optical parameters are almost identical to the ones used in Ref. \cite{liu}. The geometrical parameters are specified in Fig. 1. The permittivity of the dielectric is described by a Lorentz model \cite{note}, $
\epsilon_{d}(\omega)=\epsilon_{\infty}\left[1+\omega_{p}^{2}/(\omega^{2}_{0}-\omega^{2}-i\omega\gamma)\right]$
with $\epsilon_{\infty}=2.44$, $\omega_{p}=93.77$ THz, $\gamma=173.73$ THz and $\omega_{0}=3.1$ THz, which results in a permittivity $\epsilon_{d}\approx2.28+0.091i$, approximately constant in the relevant frequency regime, and almost identical to the one used in Ref. \cite{liu}. It should be emphasized that this specific Lorentz model for $\epsilon_{d}$ does not alter the underlying physics of MPA. The permittivity of the metal is described by a Drude model, $\epsilon_{m}(\omega)=1-\omega_{pm}^{2}/(\omega^{2}+i\omega\gamma_{m})$, with $\omega_{pm}=1.37\times10^{4}$ THz and $\gamma_{m}=40.8$ THz. Using a finite-difference time-domain method \cite{taflove}, where the size of spatial grid cell is fixed at 5 nm, we calculate the linear spectra at normal incidence and plot the results in Fig. 1. Around a wavelength of 5.93 $\mu$m, the absorption is found to be nearly 100\%. Note that this wavelength is bigger than the lattice constant $d=2\:\mu$m, so that only zero-order Bragg wave propagates to the far field. We further calculate the current function $g(z)$ at this wavelength and plot the result in Fig. 2(a). To check our numerical results, we computed the integrations in Eqs. (\ref{eq4}) and (\ref{eq5}), and verified that they are nearly identically satisfied. As discussed above, we find that the polarization current are strongly localized inside the two metallic layers, and a phase jump of $0.94\pi$ appears at the spacer-ground interface. Furthermore, we obtain $G_{c}Z_{0}=(-0.15-6.77i)$ and $G_{g}Z_{0}=(1.1+6.78i)$, in perfect agreement with the discussions above.

\begin{figure}[t]
\centering
\includegraphics[width=0.8\textwidth]{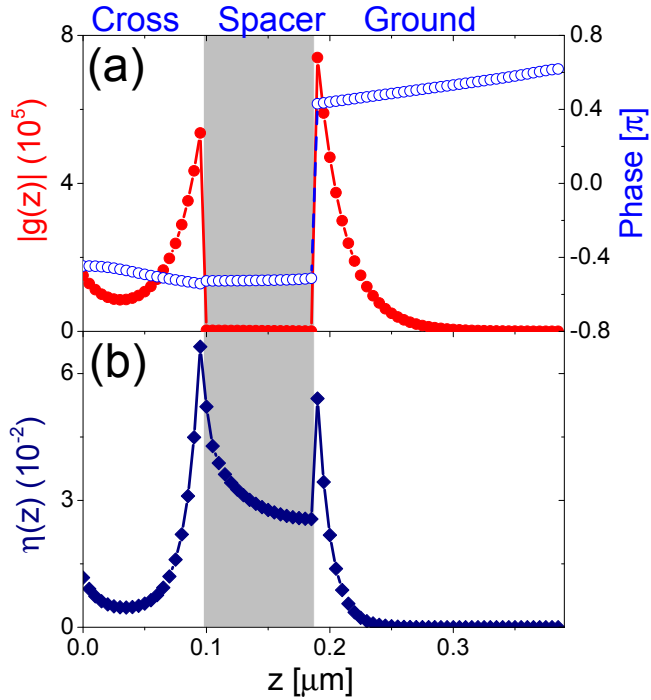}\vspace*{-4.0cm}
\caption{(a) Amplitude and phase of the function $g(z)$, and (b) the function $\eta(z)$, when $\lambda=5.93\mu$m. The dielectric layer is highlighted.} \label{fig2}
\end{figure}
\subsection{Role of non-zero-order Bragg waves}
Next we consider the dissipation of the electromagnetic energy within the MPA. According to Poynting's theorem, the energy absorption can be described by $\mathbf{J}\cdot\mathbf{E}$ \cite{jackson}. Consequently we define
\begin{equation}
\eta(z)=\frac{\int\int\textrm{Im}(\epsilon)|\mathbf{E}(x,y,z)|^{2}dxdy}{\int\int\int\textrm{Im}(\epsilon)|\mathbf{E}(x,y,z)|^{2}dxdydz}
\label{eq7}
\end{equation}
to measure the relative contribution from a specific $z$ plane. Here $\epsilon=\epsilon(\mathbf{r},\omega)$ describes the permittivity of the whole MPA structure. In general, the electric field intensity inside a metamaterial is highly inhomogeneous, which is an indication of the appearance of high-order Bragg waves, since the zero-order mode only gives a homogeneous field distribution. Therefore one can conclude that high-order Bragg waves definitely consume electromagnetic energy inside any nanostructure, in particular a MPA.

We numerically calculate $\eta$ for the structure shown in Fig. 1, and plot the result in Fig. 2(b). Under normal incidence, the two approximate theories proposed in Ref.\cite{landy} and Ref. \cite{chen} consider only the zero-order mode ($m=n=0$), which is a plane wave propagating along the $z$ direction. The corresponding EM field of the $m=n=0$ mode is parallel to the interface and must be continuous across the spacer-ground interface. If only this mode consumes electromagnetic energy, we expect that $\eta$ of the spacer layer is much smaller than that of the ground layer because $\textrm{Im}(\epsilon_{m})/\textrm{Im}(\epsilon_{d})\approx 3000$, which however does not agree with our full-wave numerical result (a similar calculation was reported in Ref. \cite{liu}). This disagreement implies that higher-order Bragg waves (contained in full-wave simulation but omitted in effective medium theory) contribute significantly to the dissipation of energy within the metamaterial. Indeed, we find numerically that the ratio of $|E_{z}/E|$ inside the spacer layer ranges from 0.84 to nearly 1.0, suggesting that non-zero-order waves have a dominant contribution to the total field in the spacer region. As discussed in the previous paragraph, one can alternatively infer the existence of high-order Bragg waves from the highly localized electric field distribution inside the MPA, which is shown in Fig. 4 of Ref.\cite{liu}.

\section{Conclusions}

To sum up, we studied the metamaterial perfect absorber in the framework of grating theory. We proved analytically that there always exists one circulating current loop (akin to a magnetic dipole), together with an electric dipole as well as an electric quadrupole. We further showed that non-zero-order Bragg waves may contribute significantly to the dissipation of the electromagnetic energy inside the perfect absorber. Such an understanding of the process in the microscopic scale is important in exploring potential applications of metamaterial absorbers.


We acknowledge support from the LANL LDRD program. This work was carried out under the auspices of the National Nuclear Security Administration of the U.S. Department of Energy at Los Alamos National Laboratory under Contract No. DE-AC52-06NA25396.


\begin{thebibliography}{99}
\bibitem{note} Because our finite-difference time-domain approach cannot handle a permittivity with a nondispersive imaginary part, we adapt a dispersive Lorentz model for the dielectric.
\end{thebibliography}
\end{document}